\documentstyle[sprocl,epsf]{article}

\begin{document}
\title{Review of $N N$ Interaction from Lattice QCD}

\author{David Richards}

\address{Dept.\ of Physics, Old Dominion University, Norfolk, VA
23529, USA\\
Jefferson Laboratory, Newport News, VA 23606, USA\\E-mail:
dgr@jlab.org} 


\maketitle

\abstracts{
Historically, lattice studies of QCD have concentrated on
understanding the properties and
structure of an isolated hadron.  Recently, there have efforts at
understanding the interactions between hadrons.  In
this talk I will review two approaches to this problem.}

Understanding the strong interaction in multi-hadron systems from
lattice QCD is a notoriously difficult problem.  Multi-hadron states
involve the computation of a four-point function and are relatively
massive, and therefore the corresponding correlation functions quickly
vanish into noise.  Furthermore,
they are large, and therefore the spatial extent of
the lattice needs to be correspondingly larger than that used in
simple spectroscopy.  Finally, the use of a Euclidean lattice
obscures the the extraction of the phase information of the full
scattering matrix.\cite{maiani90} Despite these difficulties, the
problem is so fundamental that two approaches have been adopted.

\begin{figure}[t]
\begin{center}
\epsfxsize=250pt\epsfbox{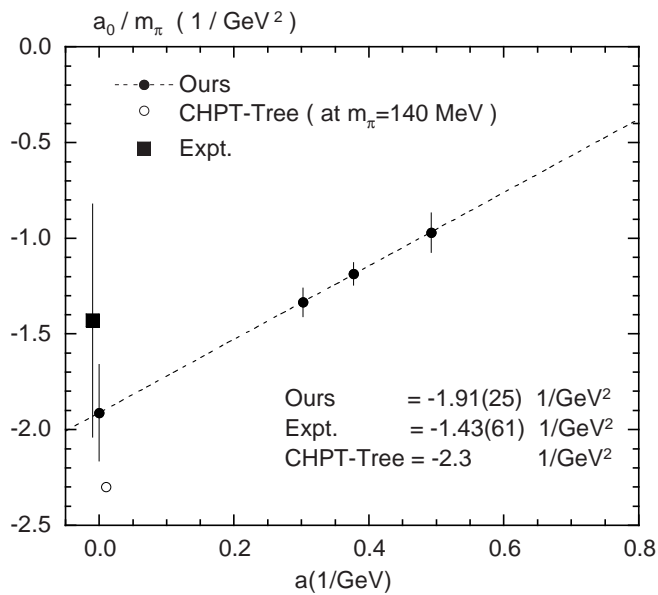}
\end{center}
\vspace{-0.3cm}
\caption{The $I=2$ pion scattering length obtained in the
quenched approximation to QCD is shown as a function of the lattice
spacing, and in the continuum limit.\protect\cite{Fuk99}  Also shown
as the open circle is the current-algebra prediction for $m_{\pi} =
140\,\mbox{MeV}$.}\label{fig:swave_pion}
\vspace{-0.4cm}
\end{figure}

L\"{u}scher demonstrated how to exploit the finite-volume shift of the
discrete two-particle energies to obtain infinite volume $s$-wave
scattering lengths.\cite{Lue86,Lue91} The application of the method to
the $\pi-\pi$ system has been encouraging,\cite{Fuk95,Fuk99} and the
$I = 2$ $\pi\pi$ scattering length is shown in
Figure~\ref{fig:swave_pion}.  

The investigation of the $NN$ system, where scattering lengths are
around $10~\mbox{fm}$, is more problematical.  Therefore the second
approach realizes that information about the nucleon-nucleon
interaction can be gleaned from the study of a simpler system, two
heavy-light mesons with static heavy
quarks.\cite{Ric89,Mih97,koniuk98,Mic99} This exhibits many of the
features of the nucleon-nucleon system, such as quark exchange,
flavor exchange and color polarization, but also admits a relative
coordinate and hence an adiabatic potential.  

The first study considered only the flavor-exchange
process,\cite{Ric89} and showed that the interaction could be
described in terms of meson exchange, with the exchanged particle
being a vector ($V$) or pseudoscalar ($P$ ) for the interaction $P P
\rightarrow PP$ or $PV \rightarrow VP$ respectively.  The study of
non-flavor-exchange contributions requires the lattice evaluation of
``all-to-all'' quark propagators.  Recently, techniques for computing
these have been developed, and evidence sought for nuclear binding
from studies of the potential.\cite{Mih97,koniuk98,Mic99} This is a
burgeoning area of research in the lattice community, and an crucial
element of the research program of the \textit{Lattice Hadron Physics
Collaboration.}

This work was supported by DOE contract DE-AC05-84ER40150 under which
SURA operates the Thomas Jefferson National Accelerator Facility.

\end{document}